\documentclass[12pt]{iopart}
\newcommand{\BEQ}{\begin{equation}}     
\newcommand{\BEA}{\begin{eqnarray}}
\newcommand{\EEQ}{\end{equation}}       
\newcommand{\EEA}{\end{eqnarray}}

\begin{document}

\input epsf.sty

\title{Kinetic roughening, global quantities, and fluctuation-dissipation relations}

\author{Yen-Liang Chou$^1$ and Michel Pleimling$^1$}
\address{$^1$Department of Physics, Virginia Tech, Blacksburg, Virginia
24061-0435, USA}
\eads{\mailto{ylchou@vt.edu}, \mailto{Michel.Pleimling@vt.edu}}

\begin{abstract}
Growth processes and interface fluctuations can be studied through the properties of global quantities.
We here discuss a global quantity that not only captures better the roughness of an interface than the
widely studied surface width, but that is also directly conjugate to an experimentally accessible
parameter, thereby allowing us to study in a consistent way the global
response of the system to a global change of external conditions. Exploiting the full analyticity
of the linear Edwards-Wilkinson and Mullins-Herring equations, we study in detail various two-time
functions related to that quantity. This quantity 
fulfills the fluctuation-dissipation theorem when considering steady-state equilibrium fluctuations.
\end{abstract}
\maketitle

\section{Introduction}
Due to its omnipresence in many fields in physics and engineering, kinetic roughening
has attracted much attention over the years, see
\cite{Mea93,Hal95,Bar95,Kru97,Kru95} for some earlier reviews. 
In many instances the properties of growth processes can be understood by analysing rather simple Langevin equations.
These Langevin equations, which can be either linear or non-linear, provide the theorists with a class of systems that they can  
study in a systematic way, using a variety of techniques.
In addition, these Langevin equations have a 
wide range of applications, ranging from different growth processes to
equilibrium step fluctuations at the surface of a crystal \cite{Gie01,Dou04,Dou05,Bon05,Bus11} 
and kinetic smoothening of interfaces \cite{Ngu10}.

The simplest Langevin equations studied in this context are the linear Edwards-Wilkinson (EW) \cite{Edw82}
\begin{equation}
\frac{\partial h(\mathbf{x},t)}{\partial t}= \nu\nabla^2 h(\mathbf{x},t)+\eta(\mathbf{x},t)
\label{eqEW}
\end{equation}
and noisy Mullins-Herring (MH) \cite{Mul63} equations
\begin{equation}
\frac{\partial h(\mathbf{x},t)}{\partial t}= -\nu\nabla^4 h(\mathbf{x},t)+\eta(\mathbf{x},t),
\label{eqMH}
\end{equation}
where the noise is usually assumed to have zero mean and to be uncorrelated:
\begin{equation}
\langle \eta(\mathbf{x},t) \rangle = 0 \qquad ; \qquad \langle \eta(\mathbf{x},t) \eta(\mathbf{x}',t')\rangle=D \delta^d(\mathbf{x}-\mathbf{x}') \delta(t-t')~.
\label{eqC2eta}
\end{equation}
Here $\mathbf{x}$ is a $d$-dimensional position vector located on the surface of the substrate, whereas for growth processes
$h(\mathbf{x},t)$ is the height of a column at position $\mathbf{x}$ at time $t$.\footnote{In the following we 
mostly use the language of growth processes,
but due to the different physical situations described by the same Langevin equations our results have a broader range
of applications.}

These equations, which depend on the two parameters $\nu$ (for the EW equation $\nu$ is the surface tension or elastic constant)
and $D$ (the noise strength), are used to describe deposition with different relaxation mechanisms. For the EW equation
the surface current responsible for the smoothening can be viewed to be due to
the gravitational potential, whereas for the MH equation a surface current arises because
of a chemical potential difference. A standard way to analyse kinetic roughening
is to study the surface width, which typically 
displays the following three regimes.
At the early stages the surface
grows in an uncorrelated way during a regime that is sometimes called the random deposition regime. At a first crossover time
$t_1$, correlated growth sets in.  This correlated regime continues until a second crossover time $t_2$ at which the system
enters the steady state regime. This time $t_2$ depends on the size of the system and shifts to larger
values when increasing the linear extend of the substrate. For an infinitely large substrate $t_2$ diverges and
the system never reaches the steady state.

Quite some attention was paid recently to the ageing processes \cite{Hen10} that take place during the
correlated growth regime \cite{Kal99,Rot06,Rot07,Bus07a,Bus07b,Cho09,Cho10,Daq11,Noh10}.
Most of the studies focused on local quantities
as for example the height-height correlation function, the response of the height to a local perturbation,
the two-time roughness or the two-time incoherent scattering function \cite{Kal99,Rot06,Rot07,Bus07a,Bus07b,Cho09}.
In \cite{Cho10} we discussed 
the correlation function of the squared width and the response of the squared width to a global perturbation.
Studying changes both in $\nu$  and $D$, our work revealed that the global response 
of the surface width
depends on how the system is perturbed. In addition, we showed that in the correlated regime
the limit value of the corresponding fluctuation-dissipation
ratio only yields the trivial value zero, due to the fact that the
studied quantity, the square of the surface width, is not conjugate to any of the system parameters that are changed
in our protocols.

In this paper we propose to study a global quantity directly related to the 
effective Hamiltonian ($m$ being an even number)
\begin{equation}
H_m=\frac{\nu}{2}\int d^dx\left(\nabla^{m/2} h\right)^2
\label{eqEHamiltonian}
\end{equation}
showing up in the Hamiltonian description that yields the corresponding stochastic equation of motion:
\begin{equation}
\frac{\partial h(\mathbf{x},t)}{\partial t}=-\frac{\delta H_m}{\delta h(\mathbf{x},t)}+\eta(\mathbf{x},t).
\label{eqLa1}
\end{equation}
Inserting (\ref{eqEHamiltonian}) into (\ref{eqLa1}) yields the linear Langevin equation
\begin{equation}
\frac{\partial h(\mathbf{x},t)}{\partial t}=-\nu(i\nabla)^m h(\mathbf{x},t)+\eta(\mathbf{x},t),
\label{eqLa}
\end{equation}
where for $m=2$ we recover the EW equation, whereas $m=4$
yields the MH equation.

As we discuss in the following, our quantity, which can be computed exactly for linear Langevin equations,
has many advantages.
On the one hand, it better describes the roughness of a surface than the surface width itself. On the other
hand, as this quantity is conjugate to $\nu$, a change in $\nu$ yields a global response 
that allows us to study the corresponding fluctuation-dissipation ratio  \cite{Cug97,Cri03}. For the case of an equilibrium steady state
(as it is for example encountered for step fluctuations on crystal surfaces)
the celebrated fluctuation-dissipation theorem is recovered, something that was not the case for the surface width \cite{Cho10}.

The paper is organized in the following way. In Section 2 we introduce our quantity that we calculate exactly for
the linear Langevin equations used in the context of non-equilibrum growth and related problems. Section 3 is 
devoted to the corresponding correlation and response functions. We thereby show that we recover the fluctuation-dissipation
ratio for equilibrium steady states. 
Section 4 gives our conclusions.

\section{Global quantity conjugate to $\nu$}

Inspection of the effective Hamiltonian (\ref{eqEHamiltonian}) allows us to define both for the Edwards-Wilkinson and the
Mullins-Herring equations the following time-dependent global quantity
that is conjugate to $\nu$:
\begin{equation}
G_m(t) = \frac{1}{2} \int d^dx\left(\nabla^{m/2} h\right)^2~,
\label{eqGm}
\end{equation}
For $m=2$ the quantity $G_m$ is of course readily identified
with the total "kinetic energy."
In the following we consider as substrate $d$-dimensional lattices with linear extend $L$. For $m=2$
our quantity can then be written as
\begin{equation}
G_2(t)=\frac{1}{2}\sum_\mathbf{x}\sum_{i=1}^d\Big[h(\mathbf{x}+\mathbf{a}_i,t)-h(\mathbf{x},t)\Big]^2
\label{eqG2defined}
\end{equation}
whereas for $m=4$ we obtain:
\begin{equation}
G_4=\frac{1}{2}\sum_\mathbf{x}\left\{\sum_{i=1}^d
\Big[h(\mathbf{x}+\mathbf{a}_i,t)-2h(\mathbf{x},t)+h(\mathbf{x}-\mathbf{a}_i,t)\Big]\right\}^2~.
\label{eqG4defined}
\end{equation}
The vector $\mathbf{x}$ now labels all points of the substrate lattice, whereas the vectors $\mathbf{a}_i$, $i = 1, \cdots ,d$,
indicate the primitive vectors on the substrate.
For simplicity the lattice constant is assumed to be unity. 
Note that geometrically $G_2$ is the total surface slope \footnote{A generalisation of this quantity to the height difference 
between columns separated by a displacement $\mathbf{r}$ has been called the second order height difference 
correlation function \cite{Kru97}.} and $G_4$ is the sum of the curvatures.

\begin{figure}[h]
\centerline{\epsfxsize=5.25in\ \epsfbox{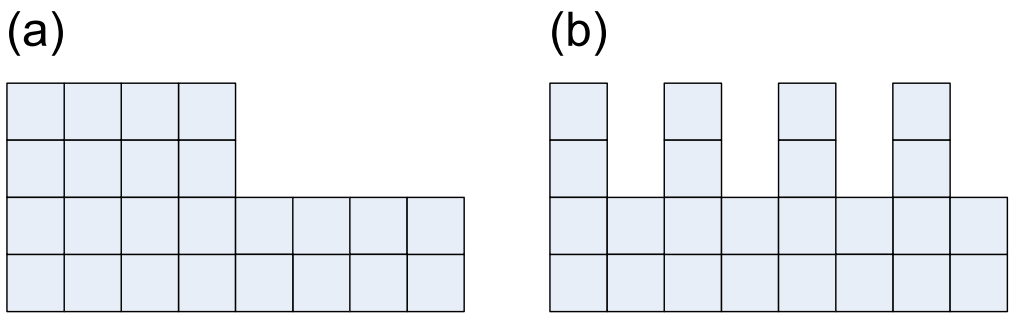}}
\caption{Two surfaces with the same surface width but different values of $G_m$.}
\label{fig1}
\end{figure}

Usually the roughness of a growing interface of columns of hight $h(\mathbf{x})$ at substrate site $\mathbf{x}$ 
is expressed by the surface width
\begin{equation}
W(t) = \sqrt{ \frac{1}{L^d} \sum\limits_{\mathbf{x}} \left( h(\mathbf{x},t) - \overline{h}(t) \right)^2}
\end{equation}
where $\overline{h}(t) = \frac{1}{L^d} \sum\limits_{\mathbf{x}} h(\mathbf{x},t)$ is the average height at time $t$.
The quantity $G_m$ can also be used to quantify the roughness of a surface \footnote{See \cite{Hua10} for
an example where the surface slope $G_2$ has been analysed in a numerical study}. In fact, it can even be argued that
$G_m$ captures the roughness much better than $W$. To see this, consider the two surfaces shown in Fig. \ref{fig1}
which have the same number of deposited particles. Whereas intuitively we would judge the surface (b) to be rougher
than the surface (a), the surface width yields the same value for both cases. The value of $G_m$, on the other
hand is larger for surface (b) than for surface (a), and this for both cases $m = 2$ and $m =4$. 

For our purpose we need to derive the exact expressions for the average of $G_m$ from the solution of the corresponding
linear Langevin equations. Writing both the height $h(\mathbf{x},t)$ and the noise $\eta(\mathbf{x},t)$ as a sum over reciprocal lattice
vectors, we obtain the following expression
\begin{equation}
\langle G_m\rangle =2^{\frac{m-2}{2}}L^d\sum_{\mathbf{q}\neq 0}\Big\langle h_\mathbf{q}h_\mathbf{-q}\Big\rangle \mathcal{P}(\mathbf{q})^{m/2},
\label{eqGm1}
\end{equation}
where
\begin{equation}
\mathcal{P}(\mathbf{q})=\sum_{i=1}^d\Big(1-\cos(q_i)\Big).
\label{eqPfactor}
\end{equation}
Here $\langle \cdots \rangle$ indicates an average over the noise. The averaged quantity $\langle G_m\rangle$
is therefore a sum over two-point correlation functions weighted by the factor $\mathcal{P}(\mathbf{q})^{m/2}$.
Using the expression \cite{Cho10}
\begin{equation} \label{eqCorrelation2hq}
\langle h_{\mathbf{q}}(t_1) h_{\mathbf{p}}(t_2) \rangle = \frac{D}{L^d\nu}e^{-\nu(q^m t_1+p^m t_2)}\frac{1}{q^m+p^m}\left(e^{\nu(q^m+p^m)t_{<}}-1\right) \delta^d_{\mathbf{q}+\mathbf{p}}
\end{equation}
for the two-point correlation function, with $q = | \mathbf{q} |$, $p = | \mathbf{p} |$, and 
$t_<$ being the smaller of the times $t_1$ and $t_2$, we obtain
\begin{equation}
\langle G_m\rangle =2^{\frac{m-4}{2}}\frac{D}{\nu}\sum_{\mathbf{q}\neq 0} \frac{1}{q^m} \left(1-e^{-2q^m\nu t}\right) \mathcal{P}(\mathbf{q})^{m/2}.
\label{eqGm2}
\end{equation}
The behaviour of $\langle G_m\rangle$ is therefore controlled by the length scale $l_t \equiv (2\nu t)^{1/m}$, similar to the 
surface width \cite{Cho10}. Depending on the relation of $l_t$ to the maximum
and minimum values of $q$, $q_{max} = \pi \sqrt{d}$ and $q_{min} = 2 \pi/L$, different regimes can be discussed.

1. For $l_t<1/q_{max}$, the system is in the random deposition regime. An expansion in small $l_t$ yields the expression
\begin{eqnarray}
\langle G_m\rangle &\approx& 2^{\frac{m-2}{2}}D t \sum_{\mathbf{q}\neq 0}\mathcal{P}(\mathbf{q})^{m/2}\nonumber\\
&\approx& 2^{\frac{m-2}{2}}D t \left(\frac{L}{\pi}\right)^d \int_0^\pi d\mathbf{q}\mathcal{P}(\mathbf{q})^{m/2}
\end{eqnarray} 
such that
\begin{equation}
\langle G_2\rangle \approx dL^dD t,
\label{eqG2L} 
\end{equation}
and
\begin{equation}
\langle G_4\rangle \approx d(2d+1)L^dD t.
\label{eqG4L} 
\end{equation}
In this regime $\langle G_m\rangle$ varies linearly in time and shows the same time dependence as the
squared surface width \cite{Cho10}. 

2. When $l_t>1/q_{min}$, the system is in the saturation regime and $\langle G_m\rangle$ reaches its maximal value,
\begin{eqnarray}
\langle G_m\rangle =G_{s,m}&\approx& 2^{\frac{m-4}{2}}\frac{D}{\nu} \sum_{\mathbf{q}\neq 0}\frac{1}{q^m}\mathcal{P}(\mathbf{q})^{m/2}\nonumber\\
&\approx& 2^{\frac{m-4}{2}}\frac{D}{\nu}  \left(\frac{L}{\pi}\right)^d I(m,d),
\end{eqnarray}
with $I(m,d)\equiv \int_0^\pi \mathcal{P}(\mathbf{q})^{m/2}/q^m d\mathbf{q}$. The numerically evaluated values of $I(m,d)$ for various $d$ and $m$ are shown in Table (\ref{tableImd}).

\begin{table}[tbp]
\begin{center}
\begin{tabular}{c c c c c}
\hline
    & d=1 & d=2 & d=3 & d=4 \\ \hline\hline
m=2 & 1.21532 & 3.43997 & 10.29687 & 31.53744 \\ \hline
m=4 & 0.49733 & 1.26083 & 3.55376 & 10.51476 \\ \hline
\end{tabular}
\end{center}
\caption{ The numerically evaluated values of  $I(m,d)\equiv \int_0^\pi \mathcal{P}(\mathbf{q})^{m/2}/q^m d\mathbf{q}$}
\label{tableImd}
\end{table}

3. To calculate the asymptotic behavior in the correlated regime, where $1/q_{max}< l_t < 1/q_{min}$, 
we rewrite Eq. (\ref{eqGm2}) in the form
\begin{equation}
\langle G_m\rangle = G_{s,m} - 2^{\frac{m-4}{2}}\frac{D}{\nu}\sum_{\mathbf{q}\neq 0} \frac{1}{q^m} e^{-2q^m\nu t} \mathcal{P}(\mathbf{q})^{m/2}
\label{eqGm5}
\end{equation}
and take advantage of the hyperspherical symmetry
of $\mathcal{P}(\mathbf{q})$ for $q<1$: $\mathcal{P}(\mathbf{q}) \approx q^2/2$. 
In fact, since (\ref{eqGm5}) involves the factor $e^{-q^ml_t^m}$, we can neglect contributions with $q>1$,
provided that $l_t \gg 1$. In that case we can simply replace the sum by an integral, yielding the result
\begin{eqnarray}
\langle G_m\rangle &\approx& G_{s,m}-\frac{D}{4\nu}\sum_{\mathbf{q}\neq 0}e^{-2q^m\nu t}\nonumber\\
&\approx&  G_{s,m}-\frac{D}{4\nu}\left(\frac{L}{2\pi}\right)^d\Omega_d\frac{\Gamma{(\frac{d}{m})}}{m}\left(2\nu t\right)^{-\frac{d}{m}}
\label{eqGmP}
\end{eqnarray}
for the correlated regime, where $\Omega_d$ is the surface of the $d-$dimensional unit sphere.

\begin{figure}
\centerline{\epsfxsize=5.25in\ \epsfbox{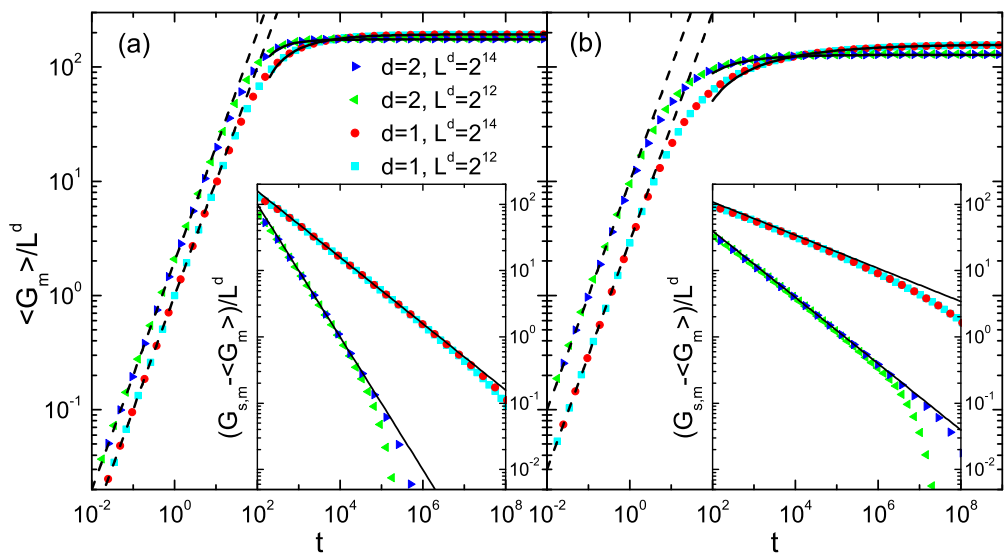}}
\caption{$\langle G_m \rangle$ for (a) the Edwards-Wilkinson case $m =2$ and (b) the Mullins-Herring case
$m=4$. Systems of various lengths and various dimensionality of the substrate are shown. 
The insets display the power-law approach to the stationary value $G_{s,m}$. The symbols are obtained by numerically 
evaluating the exact expression Eq. (\ref{eqGm2}). The dashed and solid lines are given by the asymptotic 
expressions (\ref{eqG2L},\ref{eqG4L}) and (\ref{eqGmP}). The system parameters are $\nu=0.001$ and $D=1$.
}
\label{fig2}
\end{figure}

Figure \ref{fig2} shows $\langle G_m\rangle$, obtained by numerically
evaluating Eq. (\ref{eqGm2}), for various values of $m$ and $d$ and various system sizes.
Also shown as lines are the asymptotic expressions (\ref{eqG2L}) and (\ref{eqG4L}) (dashed lines) and (\ref{eqGmP})
(full lines). The plots and the asymptotic expressions show that $\langle G_m\rangle$ grows linearly with time $t$
when the system is in the random deposition regime, the pre-factor being independent of the value of $\nu$. 
After the system passes the first crossover point, a pre-factor that depends on $\nu$ shows up. 
In the correlated regime $\langle G_m\rangle$ approaches the steady-state value $G_{s,m}$ algebraically,
with an exponent $\frac{d}{m}$. 

We should point out that the plots of $\langle G_m\rangle$ versus time, shown in figure \ref{fig2}, do not easily
reveal the crossovers between the different growth regimes, in contrast to the surface width where the different
regimes are readily identified by mere inspection. This might somehow restrict the usefulness of $\langle G_m\rangle$,
especially in situations where the steady state value is not readily known. Still, as we discuss in the
following, the quantity $G_m$ has many other advantages that makes it appealing both for theoretical and
experimental studies.

\section{Correlation function, response function, and fluctuation-dissipation ratio}
Two-time quantities are extremely useful when studying relaxation phenomena, as they allow to capture most of the 
processes that underly the properties of a system far from its steady state. Typical quantities are the correlation
and response functions, as well as combinations of these as for example the fluctuation-dissipation ratio. 
In the context of non-equilibrium growth processes and interface fluctuations
both local \cite{Kal99,Rot06,Rot07,Bus07a,Bus07b,Cho09,Ngu10}
and global \cite{Cho10} quantities have been studied to some extent.
The former include the height-height correlation function, whereas the change of the surface width to a global 
perturbation is an example of the latter.
In the following we are deriving the corresponding exact expressions that involve the global quantity $G_m$.

We first consider the two-time correlation function of $G_m$. Starting from Eq. (\ref{eqGm1}) we have
\begin{eqnarray}
\langle G_m(t)G_m(s)\rangle &=& 2^{m-2}L^{2d}\sum_{\mathbf{q},\mathbf{p}\neq 0}\mathcal{P}(\mathbf{q})^{m/2}\mathcal{P}(\mathbf{p})^{m/2}\nonumber\\
&&\Big\{\Big\langle h_\mathbf{q}(t)h_\mathbf{-q}(t)\Big\rangle\Big\langle h_\mathbf{p}(s)h_\mathbf{-p}(s)\Big\rangle\nonumber\\
&&+ \Big\langle h_\mathbf{q}(t)h_\mathbf{p}(s)\Big\rangle\Big\langle h_\mathbf{-q}(t)h_\mathbf{-p}(s)\Big\rangle\nonumber\\ 
&&+ \Big\langle h_\mathbf{q}(t)h_\mathbf{-p}(s)\Big\rangle\Big\langle h_\mathbf{-q}(t)h_\mathbf{p}(s)\Big\rangle \Big\}. 
\end{eqnarray}
Here we note that the first term yields $\langle G_m(t)\rangle\langle G_m(s)\rangle$. Since $\mathcal{P}(\mathbf{q})$ is an even function, we have
\begin{equation}
\mathcal{P}(\mathbf{q})^{m/2}\mathcal{P}(\mathbf{-q})^{m/2}
=\mathcal{P}(\mathbf{q})^{m/2}\mathcal{P}(\mathbf{q})^{m/2}
=\mathcal{P}(\mathbf{q})^m.
\label{eqPfactor2}
\end{equation}
Using the Eqs. (\ref{eqCorrelation2hq}) and (\ref{eqPfactor2}) we obtain the connected two-point correlation function of the quantity $G_m$ :
\begin{eqnarray}
C_G(t,s)&\equiv&
\langle G_m(t)G_m(s)\rangle-\langle G_m(t)\rangle\langle G_m(s)\rangle\nonumber\\
&=&2^{m-3}\frac{D^2}{\nu^2}\sum_{\mathbf{q}\neq 0}\frac{1}{q^{2m}}e^{-2 q^m \nu t}\left(e^{2 q^m \nu s}+e^{-2 q^m \nu s}-2\right)\mathcal{P}(\mathbf{q})^m.
\label{eqGCorrelation}
\end{eqnarray}

We can also calculate the evolution of the average of $G_m$ subjected to a perturbation when we suddenly change 
$\nu$ during the growth process. This is a natural thing to do as in the effective Hamiltonian 
(\ref{eqEHamiltonian}) $\nu$ and $G_m$ are conjugate quantities. Note that this is the response of a global quantity to a global
perturbation.

In this case, the evolution of the average of $G_m$ can be written as
\begin{equation}
\langle G_m \rangle_{\mu\rightarrow \nu}(t,s)=2^{\frac{m-2}{2}}L^d\sum_{\mathbf{q}\neq 0}\Big\langle h_\mathbf{q}^{(\mu\rightarrow \nu)}h_\mathbf{-q}^{(\mu\rightarrow \nu)}\Big\rangle \mathcal{P}(\mathbf{q})^{m/2},
\label{eqGmDnu}
\end{equation}
where the notation "$\mu\rightarrow \nu$" indicates that the 
change from $\mu$ to $\nu$ at the waiting time $s$. 

As discussed in \cite{Cho10,Cho09}, the solution of the
Langevin equation at times $t > s$ becomes under that change
\begin{equation}
h^{(\mu\rightarrow \nu)}_{\mathbf{q}}(t)=e^{-\nu\mathbf{q}^{m}(t-s)}h_{\mathbf{q},\mu}(s)+\int_s^t d\tau e^{-\nu\mathbf{q}^{m}(t-\tau)}\eta_\mathbf{q}(\tau)
\label{eqSFTLa2},
\end{equation}
where
\begin{equation}
h_{\mathbf{q},\mu}(s)=\int_0^s d\tau e^{-\mu\mathbf{q}^{m}(s-\tau)}\eta_\mathbf{q}(\tau)
\end{equation}
is the solution of a surface that evolves until time $s$ at the value
$\mu$ when starting from a flat initial state.

Plugging Eq. (\ref{eqSFTLa2}) into Eq. (\ref{eqGmDnu}) and comparing with Eq. (\ref{eqGm2}), we straightforwardly obtain
the response function
\begin{eqnarray}
\chi_G(t,s)
&\equiv&\frac{\langle G_m \rangle_{\mu\rightarrow \nu}(t,s)- \langle G_m \rangle_{\nu}(t)}{\epsilon}\nonumber\\
&=& 2^{\frac{m-4}{2}}\frac{D}{\epsilon}\sum_{\mathbf{q}\neq 0}\frac{1}{q^m}e^{-2q^m\nu(t-s)}\left\{\frac{1}{\mu}\left(1-e^{-2q^m\mu s}
\right) \right. \nonumber \\
&& \left. -\frac{1}{\nu}\left(1-e^{-2q^m\nu s}\right)\right\}\mathcal{P}(\mathbf{q})^{\frac{m}{2}},
\label{eqGResponse}
\end{eqnarray}
where $\epsilon=\nu-\mu$.

With these exact expressions for the correlation and response functions, we can now discuss the properties of the two-time
quantities in the different regimes. In addition we can also study the behavior of composite quantities, as for example the
fluctuation-dissipation ratio:
\begin{equation}
X(t,s) \equiv \frac{\partial \chi_G(t,s)}{\partial s}/\frac{\partial C_G(t,s)}{\partial s}.
\label{eqFDRGS}
\end{equation}

The asymptotic steady-state behavior (i.e. the large $\nu s$ limit) is easy to obtain for both quantities.
Considering only the contribution from terms with minimum $\mathbf{q}$, we have
\begin{equation}
C_G(t,s)\approx\frac{dD^2}{4\nu^2}e^{-2 q_{min}^m \nu (t-s)}
\end{equation}
and
\begin{equation}
\chi_G(t,s)\approx\frac{dD}{2\nu^2}e^{-2 q_{min}^m \nu (t-s)},
\end{equation}
where we have applied the approximation
\begin{equation}
\mathcal{P}(\mathbf{q}_{min})/q_{min}^2\approx 1/2
\end{equation}
to both expressions and, in addition, the limit $\epsilon = \nu - \mu \rightarrow 0$ to the response function.
For the fluctuation-dissipation ratio (\ref{eqFDRGS}) we obtain in the steady state, with $t \longrightarrow \infty$,
\begin{equation}
X = \frac{2}{D}.
\end{equation}
Assuming the validity of the Einstein relation $D=2T$, which makes the steady state to be an equilibrium steady state,
we recover the fluctuation-dissipation theorem $X = 1/T$, as expected.

The most interesting case is the case where both the waiting and observation times are in the correlated regime (to 
leading order identical results are obtained when the waiting time is still in the initial RD regime). This
corresponds to $1/q_{max} < l_s, l_t < 1/q_{min}$, with $l_s < l_t$ and $l_t \gg 1$. 
In this limit, we replace in both Equations (\ref{eqGCorrelation}) and (\ref{eqGResponse}) the sums by integrals 
and treat the integrands as hyperspherical symmetric functions (this is the same method we used to derive the 
asymptotic equation for $\langle G_m\rangle$ in the correlated regime). We thereby obtain the following
power-law decay functions
\begin{equation}
C_G(t,s)\approx D^2s^2\left(\frac{L}{2\pi}\right)^d\Omega_d\frac{\Gamma{\left(2+\frac{d}{m}\right)}}{2m}(2 \nu t)^{-2-\frac{d}{m}},
\label{eqGCorrelationP}
\end{equation}
and
\begin{equation}
\chi_G(t,s)\approx Ds^2\left(\frac{L}{2\pi}\right)^d\Omega_d\frac{\Gamma{\left(2+\frac{d}{m}\right)}}{2m}\left[2 \nu (t-s)\right]^{-2-\frac{d}{m}} 
\label{eqGResponseP}
\end{equation}
where we replaced $e^{2 q^m \nu s}+e^{-2 q^m \nu s}-2$ by $(2q^m\nu s)^2$ for the correlation function and 
$(1-e^{-2q^m\mu s})/\mu-(1-e^{-2q^m\nu s})/\nu$ by $2(\nu-\mu)s^2q^{2m}$ for the response function before integrating. 
These replacements correspond to retaining only the leading terms in the Taylor expansions.

\begin{figure}
\centerline{\epsfxsize=5.25in\ \epsfbox{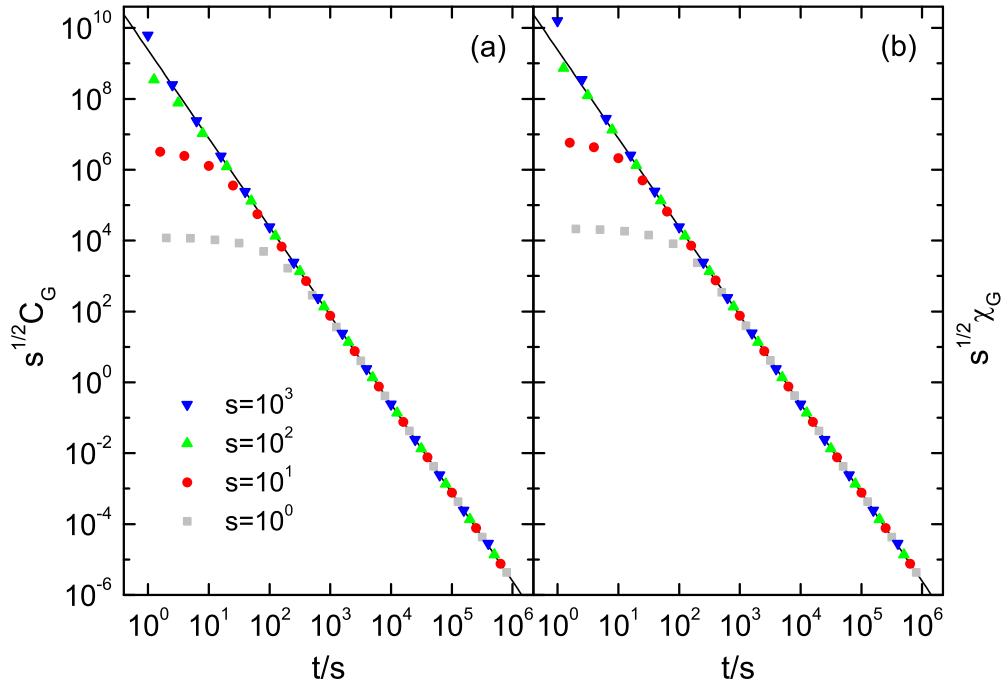}}
\caption{Ageing scaling of (a) the correlation and (b) the response function of $G_2$ in the one-dimensional EW system. 
The parameters for the calculations are $L=2^{14}$, $\nu=0.001$, and $D=1$.
}
\label{fig3}
\end{figure}

In the literature on physical ageing it is convention to write for a system undergoing simple ageing the 
two-time correlation and integrated response functions in the form \cite{Hen10}
\begin{equation}
C_G(t,s) = s^{-b} f_C(t,s) ~~~; ~~~ \chi_G(t,s) = s^{-a} f_\chi(t,s)~,
\label{eqaging}
\end{equation}
where $f_C(y)$ and $f_\chi(y)$ are scaling functions which decay algebraically for large arguments:
\begin{equation}
f_C(y) \sim y^{-\lambda_C/z} ~~~; ~~~ f_\chi(y) \sim y^{-\lambda_\chi/z}~,
\end{equation}
where $\lambda_C$ and $\lambda_\chi$ are the autocorrelation and autoresponse exponents, whereas $z$ is the dynamical
exponent (for our linear Langevin equations we have that $z = m$).
Recasting Equations (\ref{eqGCorrelationP}) and (\ref{eqGResponseP}) in these ageing forms, we immediately obtain that
$a=b=d/m$, whereas the autocorrelation and autoresponse exponents are given by $\lambda_C=\lambda_\chi=2+d/m$.
This ageing scaling is illustrated in Figure \ref{fig3} for the one-dimensional EW system.

\begin{figure}
\centerline{\epsfxsize=5.25in\ \epsfbox{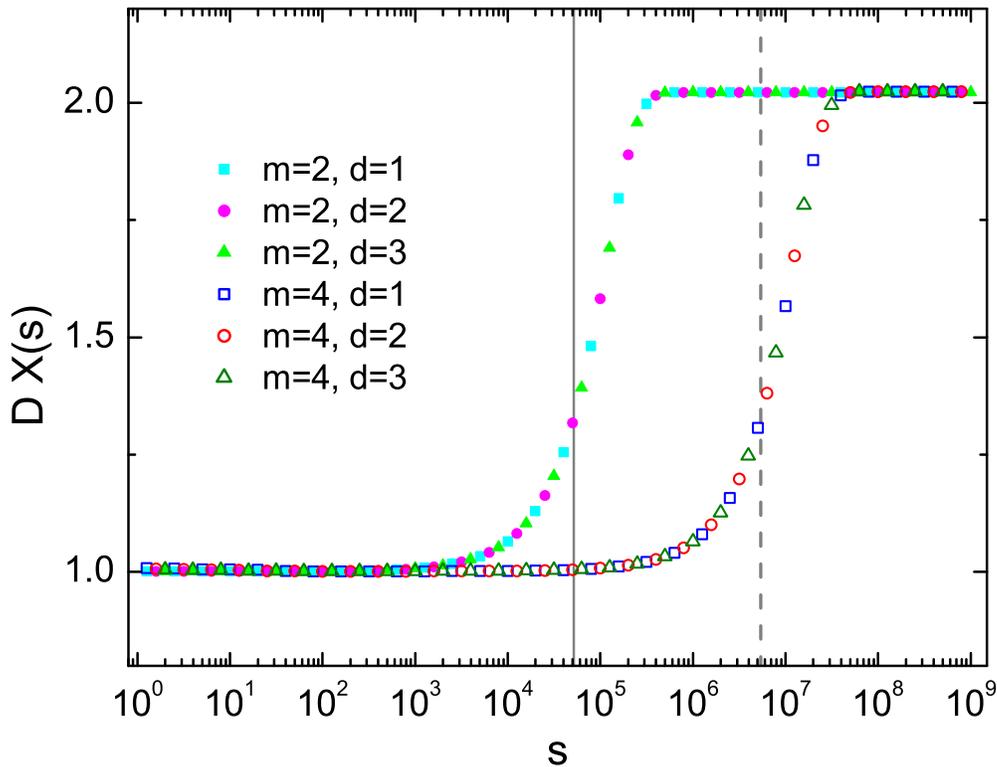}}
\caption{The fluctuation-dissipation ratio for $m=2$ and $m=4$ in various dimensions as a function of the waiting time $s$. 
Here $X(s) = X(s+10^6,s)$. The linear extension for all systems is $L=2^6$.
The full (dashed) line indicates the crossover time $t_2$ for the EW (MH) equation.
Note that the value of $X(s)$ is independent of the dimensionality of the substrate.
}
\label{fig4}
\end{figure}

\begin{figure}
\centerline{\epsfxsize=4.25in\ \epsfbox{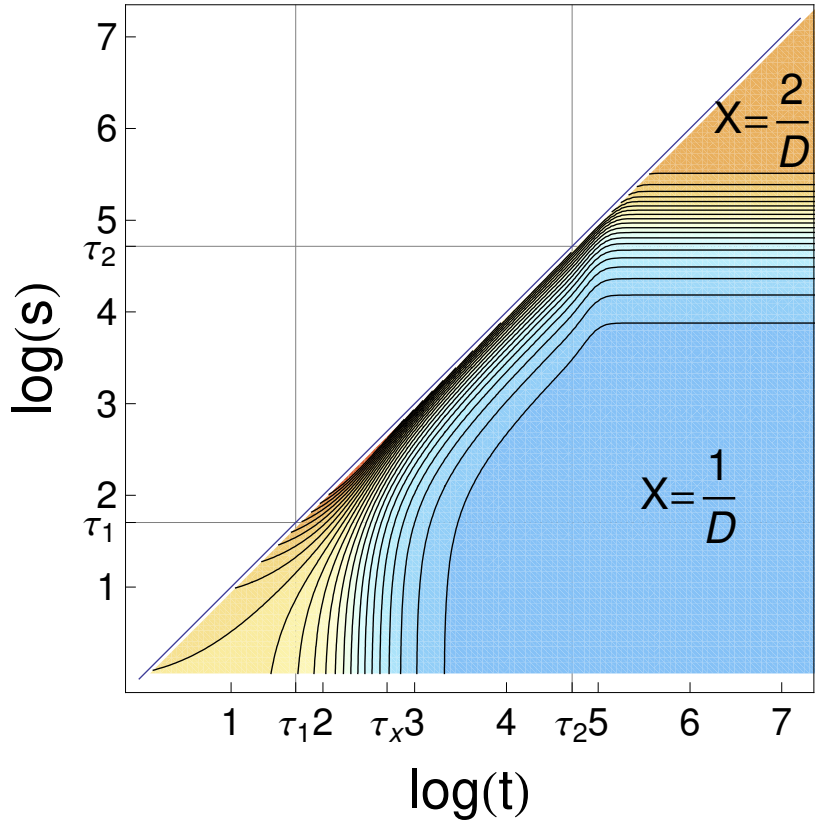}}
\caption{Contour plot of $X(t,s)$ for the one-dimensional EW equation. The parameters used here are
$L = 2^6$, $\nu = 0.001$, and $\mu = 0.99 \nu$. Also indicated are the logarithms of various relevant 
time scales: $\tau_1 = \log t_1$, $\tau_2 = \log t_2$, and $\tau_x = \log (1/2\nu)$.
Here and in the following figures, we illustrate our results for the one-dimensional EW equation, but, as the
exact results reveal, similar results are obtained for the MH equation as well as for substrates of higher
dimensionality.
}
\label{fig5}
\end{figure}

Let us close this Section with a more careful discussion of the fluctuation-dissipation ratio (\ref{eqFDRGS}).
As mentioned in the introduction, in a finite system the correlated regime goes over into the stationary state
at a system size dependent time $t_2$. In the infinite system $t_2$ diverges and the correlated regime prevails
for all times. For the limit value
\begin{equation}
X_\infty= \lim\limits_{s \longrightarrow \infty} \lim\limits_{t \longrightarrow \infty} X(t,s)
\end{equation}
we therefore obtain the values $X_\infty = 1/D = 1/2T$ for the infinite system and $X_\infty = 2/D = 1/T$ for the
finite system, where we introduced temperature via the Einstein relation. The finite system ending up in the steady
state for finite times, we recover the fluctuation-dissipation ratio. If the system remains in the correlated region,
the effective temperature is twice that of the heat-bath. The crossover between these two regimes can be visualized for
finite systems by plotting $X(s)$ for $t \gg s$ as done in Fig. \ref{fig4}. For that figure we plot the value of $X(s) =
X(s+10^6,s)$ as a function of $s$, which yields the value $X(s) = 1/D$ for $s \ll t_2$ and the value
$X(s) = 2/D$ for $s \gg t_2$. The crossover times for $m=2$ and $m=4$ are indicated in Fig. \ref{fig4} by 
the vertical lines.
Fig.\ \ref{fig5} gives a more comprehensive view of the behaviour of $X(t,s)$ as a function of both $s$ and $t$
for the one-dimensional EW equation.
Two plateaus can be distinguished in the contour plot: one for the steady state (i.e. the regime where $t > t_2$ and
$s > t_2$) where $X = 2/D$ 
and one away from stationarity, with 
$ t \gg 1/2\nu$ and $s < t_2$, where $X = 1/D$.

\section{Conclusion}

In the past the study of global quantities in systems relaxing towards a steady state has proven very fruitful in a large
variety of systems (see the corresponding discussion in \cite{Hen10}). In \cite{Cho10} we did a first attempt at using
global quantities in the context of correlated growth and interface fluctuations, 
choosing the surface width as our global quantity. However, the
surface width is a complicated quantity that has the notable drawback that the conjugate system parameter is unknown.
Consequently, it
is not possible to form a meaningful fluctuation-dissipation ratio using
that quantity.

In this paper we are proposing a different global quantity for the study of kinetic roughening and related interface
problems. This quantity is proportional to
the effective Hamiltonian used in the Langevin description and is conjugate to a system
parameter that can be changed in experiments \cite{Cho09,Ngu10}. In fact, $G_m$ seems better suited to capture the roughness
of a surface than the surface width itself, as illustrated in figure \ref{fig1}.

Focusing on linear Langevin equations we derive exact expressions for $G_m$ as well as for the corresponding correlation and
response functions. This allows us to discuss also more complicated quantities as for example the
fluctuation-dissipation ratio. In fact, we recover for the quantities derived from $G_m$
the fluctuation-dissipation theorem for equilibrium steady states, whereas in the correlated regime 
we can assign an effective temperature to our system. 

All calculations presented in this paper have
been done in the context of linear Langevin equations. However, most growth processes are governed by non-linearities.
It is therefore important to clarify to what extend our results obtained for linear Langevin equations
remain valid when considering non-linear stochastic equations as for example the Kardar-Parisi-Zhang equations \cite{Kar86}.
We intend to address this and other questions in the future.

\ack
This work was supported by the US National
Science Foundation through grant DMR-0904999.
We thank Rahul Kulkarni and Uwe C. T\"{a}uber for helpful discussions.

\end{document}